\documentclass[12pt]{article}
\usepackage{graphicx}
\begin{document}
{\bf Computer simulation of language competition by physicists}

\bigskip
\centerline{Christian Schulze and Dietrich Stauffer}

\bigskip
\centerline{Institute for Theoretical Physics, Cologne University, D-50923 
K\"oln, Euroland}

\bigskip
{\bf Abstract:}

Computer simulation of languages is an old subject, but since the
paper of Abrams and Strogatz (2003) several physics groups
independently took up this field. We shortly review their work and
bring more details on our own simulations.

\bigskip
\centerline{e-mail: stauffer@thp.uni-koeln.de}

\section{Introduction}

Human \index{languages} languages dististinguish us from other animals, but also birds
or ants have systems of communication. Also, humans have invented
alphabets and other formalized forms of writings. In principle the
methods to be described here could be applied also to these other forms of 
communication, but mostly we are interested here in the presently about $10^4$ 
different human languages on this planet\cite{grimes}. We leave it to
linguists to distinguish languages from dialects or language families;
when we mention "language" readers may read dialect or family instead.

Everyday language contains thousands of words for different aspects of
life, and with the special words of science, medicine, $\dots$ we get 
much more. For the same concept of everyday life, each different
language in general has a different word, and thus the number of
possible languages is enormous and difficult to simulate. Things
become easier if we look only at \index{grammar} grammar; do we order (subject,
object, verb) or (subject, verb, object) or $\dots$? Briscoe
\cite{briscoe} mentioned about 30 independent binary grammatical
choices, which leads to a manageable $2^{30} \simeq 10^9$ possible
languages, which can be symbolized by a string of $\ell = 30$
bits. Thus many of the simulations described here use bit-strings with 
$\ell = 8, 16 \dots 64$. 

The present situation is not in equilibrium; about every ten days a
human language dies out, and in Brazil already more than half of the 
indigenous languages have vanished as a result of the European
conquest. On the other hand, \index{Latin} Latin has split in the last two millennia 
into several languages, from Portuguese to Romanian, and many experts
believe that Latin and the other \index{Indo-European} Indo-European languages spoken 600
years ago from Iceland to Bengal (and now also in the Americas,
Australia, Africa) have originated from the people who invented
agriculture in the \index{Konya} Konya plane of Turkey, $10^4$ years ago. Thus
similar to biology, also languages can become extinct or speciate
into several daughter languages.

In contrast to biology, humans do not eat humans of other languages as
regular food, and thus one does not have a complex ecosystem of
predators eating prey as in biology. Instead, languages are meant for 
communication, and thus there is a tendency of only one language
dominating in one region, like German in Germany etc. Will
globalisation lead to all of us speaking one language in the distant
future? For physics research, that situation has already arrived many
years ago. If we follow the Bible, then at the beginning Adam and Eve
spoke one language only, and only with the destruction of the Tower of
Babel different languages originated. 

Thus in the history mankind we may have had first a rise, and later a
decay, in the number of different languages spoken. In \index{Papua} Papua New Guinea 
\cite{novotny} there are now $10^3$ languages, each spoken by about $10^3$
people; can this situation survive if television and
mobile phones become more widespread there?

While we cannot answer these questions, we can at least simulate such 
"survival of the fittest" among languages, in  a way similar but not
identical to biology. We will not emphasize here the longer history of 
computer simulations of how children learn a language \cite{nowak}, see also
\cite{roma},
or how mankind developed the very first language out of ape
sounds\cite{brighton}. Instead we talk about the competition between different
languages for adults. And we will emphasize the \index{agent-based} "agent-based" models
simulating individuals, analogous to \index{Monte Carlo} Monte Carlo and Molecular
Dynamics for spins and molecules

The second section deals with \index{differential equations} differential equations (a method we
regard as outdated), followed by agent-based simulations with few
languages in section 3 and with many languages in section 4. Further
results from the two many-language models are given in the appendix.

\section{Differential equations}

\begin{figure}[hbt]
\begin{center}
\includegraphics[angle=-90,scale=0.5]{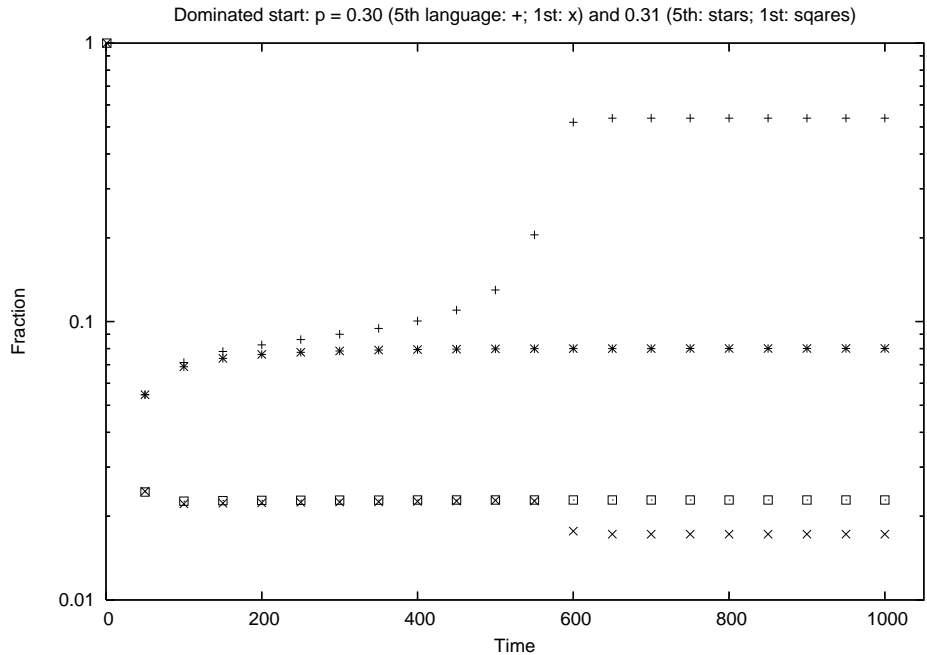}
\end{center}
\caption{
Development of one dominating language, or lack of such dominance, for
the model of Nowak et al \cite{nowak}, with random matrix
elements. We start from the dominance of another
language. The different symbols correspond to two suitably
selected languages and two slightly different mutation rates $p \simeq 0.3$.
From \cite{newbook}.
}
\end{figure}

\begin{figure}[hbt]
\begin{center}
\includegraphics[angle=-90,scale=0.5]{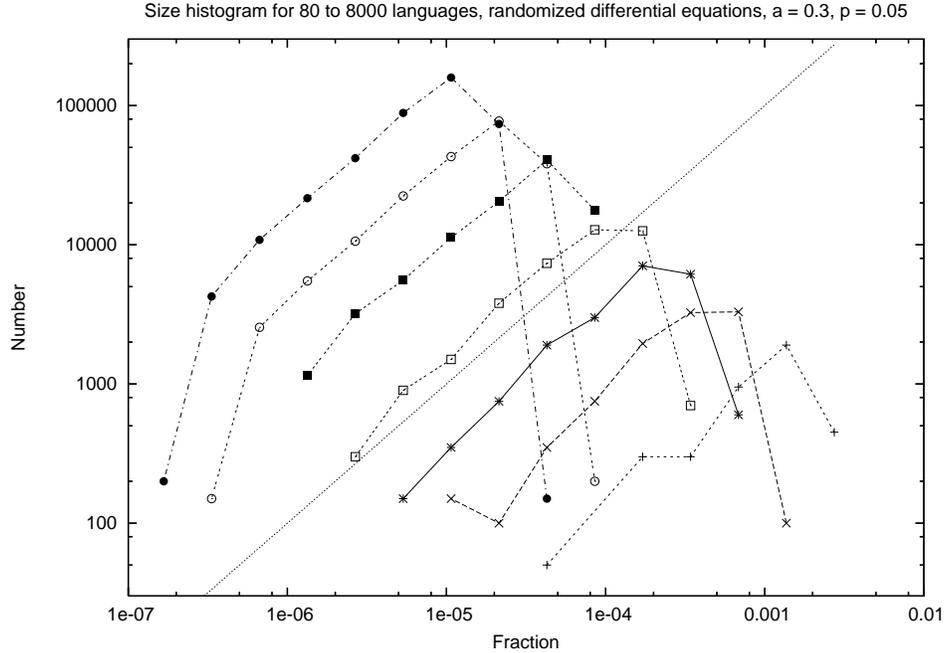}
\end{center}
\caption{Size histogram, ignoring the dominating language, for the
model of Nowak et al \cite{nowak} with random matrix
elements. The number of simulated languages varies from 80 on the
right to its real value 8000 on the left; The straight line has slope
1 in this log-log plot. From \cite{newbook}.
}
\end{figure}

\begin{figure}[hbt]
\begin{center}
\includegraphics[angle=-90,scale=0.5]{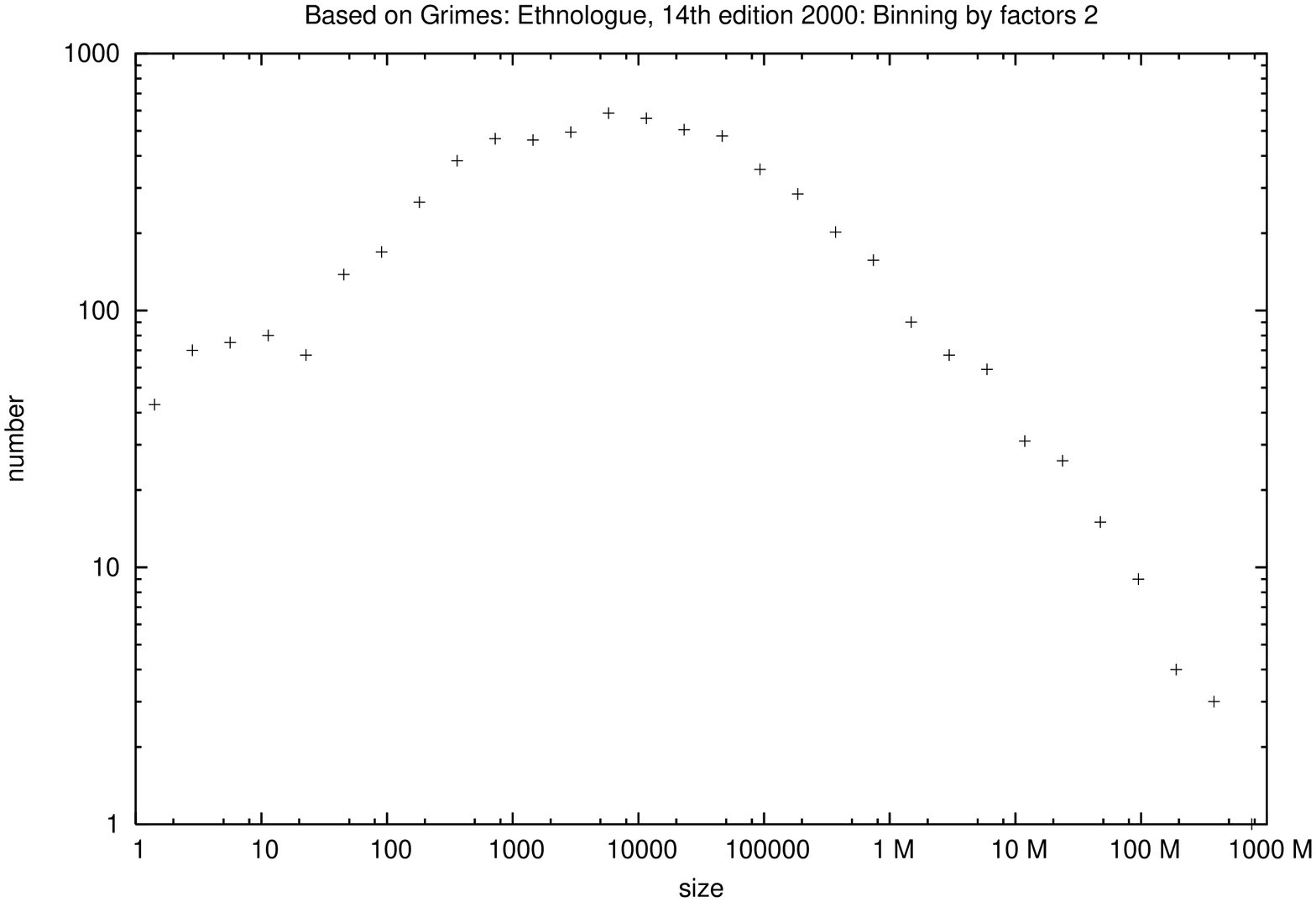}
\end{center}
\caption{Size histogram for human languages, from \cite{grimes}. 
We bin language sizes by factors of two, just
as in Fig.2: Thus the leftmost point corresponds to size = 1, the
second sums sizes 2 and 3, the third sums sizes 4 to 7, etc. 
}
\end{figure}

Already Nettle \cite{nettle} suggested a very simple differential
equation to see how the number $L$ of languages changes with time:
$$dL/dt = 70/t - L/20$$
Here the time unit is thousand years. (Actually $L$ is the number of
different language groups, and time is discrete). The second term on
the RHS means a loss of five percent per millennium; the first term
indicates the formation of new languages which became more difficult 
when the population became higher since then the higher demand for
communication reduced the chances of new languages to develop. The 
aim was to explain why the recently populated Americas have a higher
language diversity than Africa and Eurasia with their older human
population. For long times, this differential equation means that $L$
decays exponentially towards zero.

Nowak et al \cite{nowak} use
$$dx_j/dt = (\sum_i f_iQ_{ij}x_i) - \phi x_j$$
for the fraction $x_j$ of a population speaking language $j=1,2,3
\dots L$. (Actually they apply this equation to the learning of languages
or grammars by children; the interpretation for competition between
adult languages is ours.) Here the \index{fitness} fitness $f_i = \sum_j F_{ij}x_j$ of
language $i$ is determined by the degree $F_{ij}$ to which a speaker
of language $i$ is understood by people speaking language $j$. The
average fitness is $\phi = \sum_i f_ix_i$ and is subtracted to keep
the sum over all fractions $x_j$ independent of time. The
probability that children from $i$-speaking parents later speak
        language $j$ is $Q_{ij}$.

For a large number $L$ of languages, there are numerous free
parameters in the matrices $Q_{ij}$ and $F_{ij}$. With most of them the same
one finds a sharp \index{phase transition} phase transition \cite{nowak} as a function of
\index{mutation} mutation rates $Q_{ij}$. If one starts with only one
language, then at low mutation rates most of the people still speak
this language and only a minority has switched to other languages. For
increasing mutation rates, suddenly the system jumps from this \index{dominance}
dominance of one language to a \index{fragmentation} fragmentation state where 
all languages are spoken equally often. If, in turn, we start from such a
fragmented state then it stays fragmented at high mutation rates. With
decreasing mutation rates it suddenly jumps to the dominance of one
language (numerically, one then has to give this one language a very
slight advantage). The two jumps do not occur at the same mutation
rate but show hysteresis: Starting with dominance and increasing the
mutation rate allows dominance for higher mutation rates then when we
start with fragmentation and decrease the mutation rate.

Qualitatively these properties remain if the many matrix elements are
selected randomly instead of being the same \cite{newbook} except that
the hysteresis has become very small. Fig.1 shows the case where we
start with dominance and looks similar to the case where we start with
fragmentation. The time development for two of the 30 simulated
languages is shown for two slightly different mutation rates, and we
see how for the lower mutation rate but not for the higher rate one of
the two languages starts to dominate, at the expense of the other.

These 30 languages are more mathematical exercises, but Fig.2 applies
these methods to up to $L =8000$ languages, using two $8000 \times 8000$ 
random matrices $F$ and $Q$. We show the size distribution of
languages, where the size is the fraction of people speaking this
language. On this log-log plot we see roughly parabolas, shifting to
the left with increasing number $L$ of languages. These parabolas
correspond to log-normal distributions, roughly as observed
empirically in Fig.3.

(Similar to Komarova \cite{nowak} we assume the average $F$ to be 0.3 except
for $F_{ii} = 1$ and the average $Q$ to be $p/(L-1)$ except $Q_{ii}
= 1-p$; the actual values are selected randomly between zero and
twice their average.) 

There are two problems in this comparison of Figures 2 and 3: In these 
simulations, the (logarithmic) range over which the language sizes vary is 
quite small and does not change with increasing $L$. And the
real distribution is unsymmetric, having higher values for small 
languages \cite{sutherland} than the log-normal distribution; this
enhancement is missing in the simulation of Fig.2. Finally, we
cheated: Fig.2 was taken in the dominance regime and the dominating
language was ignored in the statistics.
 
Much more attractive for physicists was the one-page paper of Abrams and 
Strogatz \cite{abrams} which was within weeks followed by a poster of 
Patriarca and Leppanen \cite{finland}. This pair of papers then triggered
apparently independent research in Spain \cite{spain}, Greece \cite{kosmidis},
Germany \cite{schulze}, Argentina \cite{argentina} and at two different places
in Brazil \cite{schwammle,gomes}, all on language competition. 

\begin{figure}[hbt]
\begin{center}
\includegraphics[angle=-90,scale=0.5]{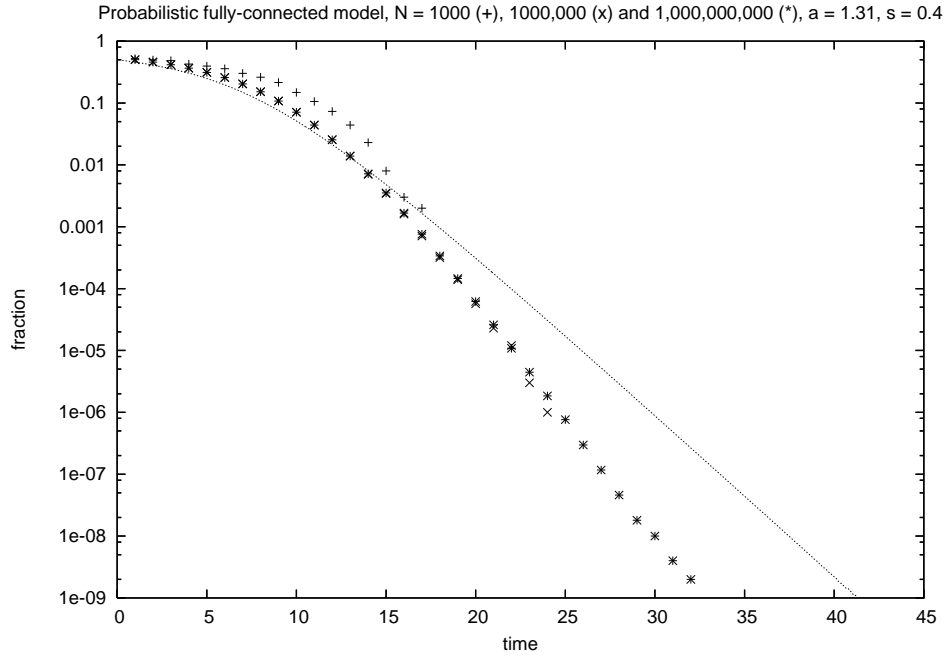}
\end{center}
\caption{ 
Exponential decay for the language with lower status, consisting 
initially of half the population. The symbols give Monte Carlo simulations 
where each individual is influenced by the whole population $N$, while the
like is the result of the differential equation of Abrams and Strogatz.
$a = 1.31, \; s = 0.4, \; N = 10^3, 10^6, 10^9$.
}
\end{figure}

\begin{figure}[hbt]
\begin{center}
\includegraphics[angle=-90,scale=0.30]{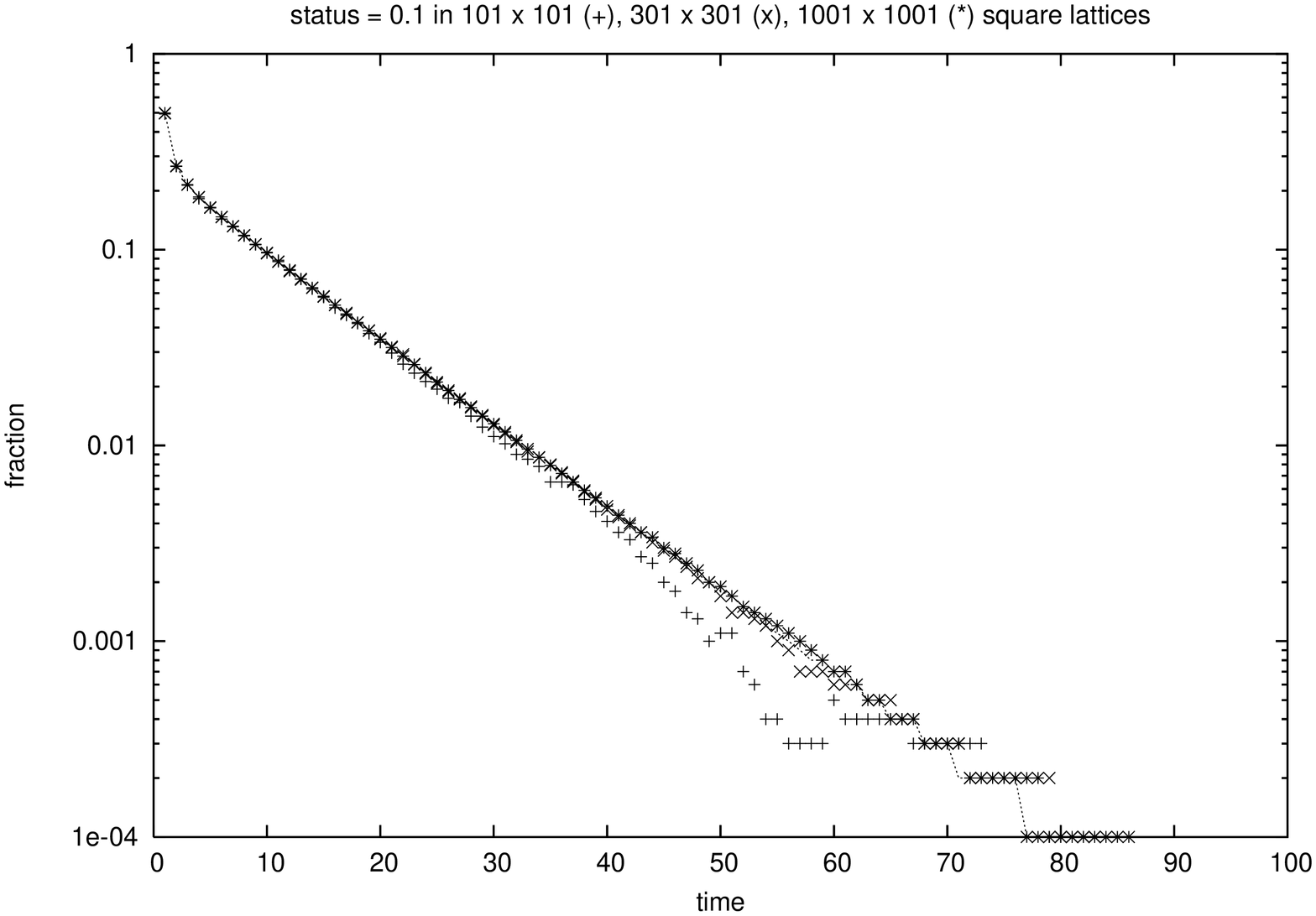}
\includegraphics[angle=-90,scale=0.30]{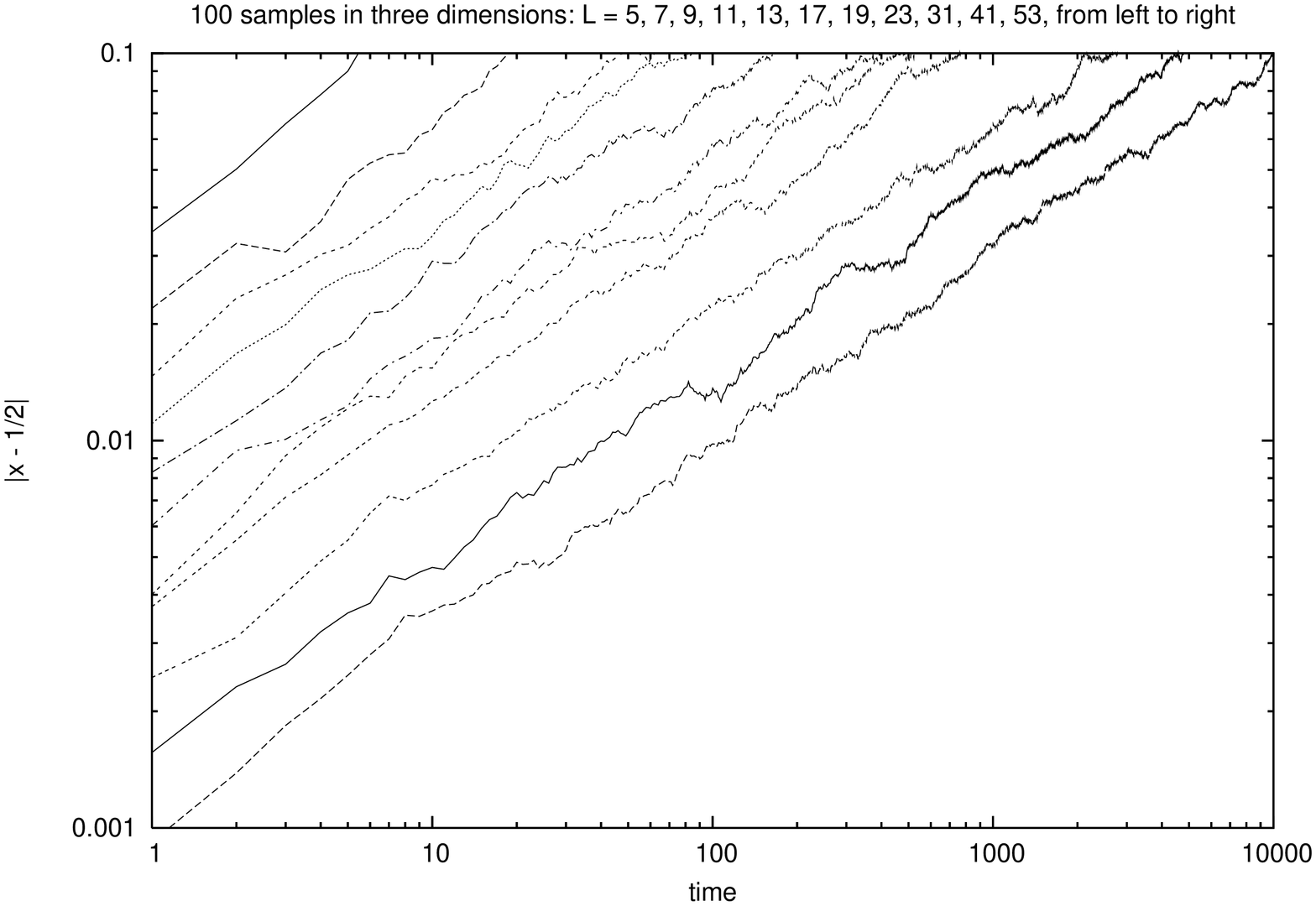}
\end{center}
\caption{ 
Part a: As for Fig.4 but on a $101 \times 101, \; 301 \times 301, \; 1001 \times 1001$ 
square lattices with $a = 1.31, \; s = 0.1$. Part b: Three-dimensional lattices,
at the symmetry point $x(t=0) = s = 1/2, \; a = 1$. After a long time, the 
concentration moves towards zero or one.
}
\end{figure}

The Abrams-Strogatz differential equation for the competition of a language Y
with higher social status $1-s$ against another language X with lower social 
status $s$ is
$$ {\rm d}x/{\rm d}t = (1-x)x^as - x(1-x)^a(1-s) $$
where $a \simeq 1.3$ and $0 < s \le 1/2$. Here $x$ is the fraction in the 
population speaking language X with lower social status $s$ while the fraction 
$1-x$ speaks language Y. Figure 4 with no status difference, $s = 1/2$, shows 
as intended that language to win which is initially in the majority; the other
language dies out. For $x(t=0) < 1/2$ the language Y wins and for $x(t=0) > 1/2$
the language X wins. This is highly plausible: If we would immigrate to Brazil
where in most places most of the people speak Portuguese, then also we would 
have to learn Portuguese, not because of status but because of numbers. If 
the initial minority language has the higher status, as happened 500 years ago
when Portuguese ships landed in Brazil, then it may win at the end, thanks to 
guns, writing, and other status aspects, as is the case in Brazil. Figures 
for unequal status are published in \cite{newbook}. 

The Finnish group \cite{finland} generalized this simple differential equation
to a square lattice, where it became a partial differential equation including
a La\-placian $\nabla^2 x({\bf r})$. Then having in one part of the lattice a 
higher status for X compared to Y, and in the other part the opposite status
relation, they showed that the languages X and Y can coexist besides each 
other, with a narrow interface in between. This reminds us of Canal Street
in \index{New Orleans} New Orleans which in earlier times separated the French quarter from 
the English speaking part. We will later return to such geographical coexistence
also without status differences, arising merely from an initial separation 
into an X part and a Y part \cite{schulze,tibihmo}.

If we set $a=1$ the simple logistic \index{Verhulst} Verhulst equation results \cite{verhulst},
$$ {\rm d}x/{\rm d}t = (2s-1)(1-x)x $$
which was applied to languages \cite{ke} already before Abrams and Strogatz.
This case was generalized to two Verhulst equations \cite{argentina} describing
the two populations of people speaking languages X and Y. Now as in 
Lotka-Volterra equations for predators eating prey, both populations can 
coexist with each other in some parameter range, which in the usual 
Abrams-Strogatz model is possible only for $x(t=0) = s = 1/2$. 

The competition of two languages is changed if some people become bilingual
 \index{bilingual}, that means they learn to speak the other language which 
was not their mother tongue \cite{spain}. This was applied to Gallego versus 
Castellano in Spain \index{Spain};
of course, some may regard Castellano spoken in Madrid as the proper Spanish, 
and Gallego as its dialect spoken in Galicia. As citizens of the Prussian 
occupied Westbank of the Rhine River, we know that publicly going into such 
details before liberation may be dangerous. A language is a dialect with an
army and a navy behind it. 

Of course, all these differential equations are dangerous approximations, just 
as \index{mean field theory} mean field theory for critical phenomena in statistical physics is 
dangerous. We know since 80 years that the one-dimensional Ising model has 
a positive Curie temperature $T_c$ in the mean-field approximation, while in 
reality $T_c = 0$. Thus do the Abrams-Strogatz results remain correct if we
deal with individuals which randomly change from one language to the other,
with probabilities corresponding to the original differential equation?

In general, the answer is yes \cite{mallorca}: As long as not both $s$ and 
the initial concentration $x(t=0)$ are 1/2, one language still dies out, and
it does so exponentially. This holds for the case of everybody influencing 
everybody, Fig.4, as well as for a square lattice where everybody is influenced
only by its four lattice neighbours, Fig.5. The line in Fig.4 is the solution
of the differential equation and agrees qualitatively with the 
\index{Monte Carlo} Monte Carlo 
results represented by the separate symbols for various total constant 
populations $N$. Only for a completely symmetric start, $s = x(t=0) = 1/2$,
when the differential equation gives an equilibrium (stable for $a < 1$ and
unstable for $a > 1$), the microscopic Monte Carlo simulation gives
one or the other language dying out, while the differential equation then
predicts both to always comprise half of the population each.
More details are given in \cite{mallorca}.

Finally we mention the model of \cite{novotny} which also does not deal with
individuals but avoids differential equations.

\section{Microscopic models}

Here we deal with the more modern methods of language simulations, based on 
individuals instead of on overall concentrations. Such methods are applied in 
physics since half a century and are called \index{agent-based} 
agent-based in some fields outside
physics. First we review two models for only two (or a few) languages, then
in much greater detail the two models for many languages. 

\subsection{Few languages}

The model of Kosmidis et al \cite{kosmidis} for mixing two languages X and Y 
uses \index{bit-string} bit-strings of length 20; each bit can be 0 (representing a word or
grammatical aspect which is not learned) or 1 (an element which this individual
has learned). If 
someone  speaks language X perfectly and language Y not at all, the bit-string
for this person is 11111111110000000000 while 00000000001111111111 corresponds
to a perfect Y-speaker. People can become perfectly bilingual, having 
all 20 bits at 1, but this is rare. This model is particularly simple to explain
the generation of a mixture language Z out of the two original languages X and 
Y. One merely has to take about ten bits equal to one and distribute them 
randomly among the 20 bit positions. This may then correspond to the creation
of Shakespeare's \index{English} English out of the Germanic language spoken by the 
Anglo-Saxons and the French spoken by the Normannic conquerors of the year 1066.

Biological \index{ageing} ageing was included in the model of Schwammle \cite{schwammle}, using
the well-established Penna model \cite{penna,verhulst,newbook} of mutation 
accumulation. Two languages X and Y are modelled. Individuals learn to speak 
from father and mother (and thus may become \index{bilingual} bilingual) and move on a square 
lattice in search of emptier regions. Bit-strings are used also here, but
only for the ageing part to store genetic diseases; the two languages have
no internal structure here.  A bilingual person surrounded by neighbours speaking
mostly language X forgets with some probability the language Y, and vice versa.
The model allows for the coexistence of the two languages, each in a different 
region of the lattice, as in \cite{finland} but without giving one language 
a higher status than the other.

In his later model \cite{schwammle}, that author allows for up to 16 languages.
Again the structure of languages is ignored.
Only young people can learn languages from others, and sometimes they learn 
a new language by themselves. As a function of the "mutation" probability to 
learn independently a new language, the model gives \index{dominance} dominance
of one language for small mutation rates, and \index{fragmentation} 
fragmentation of the population into many languages for high mutation rates, 
with a sharp \index{phase transition} phase transition separating 
these two possibilities, e.g. at a mutation rate near 1/4. This phase transition
is similar to that found by \cite{nowak} as reviewed above.

\subsection{Many languages}

To explain the existence of the $10^4$ present human languages, we need 
different models \cite{gomes,schulze} which we review now.

\subsubsection{Colonization}

After the first human beings came to the American continent by crossing the 
Bering street several ten thousand years ago, presumably they first all
spoke one language. Then they moved southward from Alaska and separated into 
different tribes which slowly evolved different languages. This first 
\index{colonization} 
colonization was modeled by Viviane de Oliveira and collaborators \cite{gomes}
by what we call the Viviane model \index{Viviane model}. 

Languages have no internal structure but are labelled by integers 1,2,3 ...
Human population starts at the centre site of a square 
lattice \index{lattice} with language
1, and from then on humans move to empty neighbour sites of already populated
areas. Each site can carry a population of up to about $10^2$ people, selected
randomly. The size or fitness of a language is the number of people  speaking
it. On every new site, the population selects as its own language that of a
populated neighbour site, with a probability proportional to the fitness of the
neighbouring language. In addition, the language can mutate into a new language
with a probability $\alpha$/size. To prevent this \index{mutation} 
mutation rate to become too
small, this denominator is replaced in their later simulations
by some maximum $\simeq 10^3$, if the actual
language size is larger than this cut-off value. The simulation stops when 
all lattice sites have been populated. A complete \index{Fortran} Fortran
program is listed in \cite{schulze}(d). 

For a mutation coefficient $\alpha = 0.256$ the simulated language sizes in the 
Viviane model can reach the thousand millions of Chinese \cite{schulze}(d),
but the shape differs from Fig.3 and corresponds more to two power laws than 
to one roughly log-normal parabola. In contrast to other models
\cite{nowak,schwammle,schulze} there is no sharp phase transition between
the dominance of one language and the fragmentation into numerous languages.
\cite{schulze}(d).

This Brazilian group \cite{brazil} earlier had claimed that
the language size distribution follows two power laws, both
indicating a decay of the number of languages with increasing 
language size. This fit, however, applies only to the large-size 
tail and not for small sizes where the power law would
indicate an unrealistic divergence. Fig.3 in contrast shows there
a very small number, with less than $10^2$ of the $10^4$ 
languages spoken by only one person \cite{sutherland}. The
cumulative number of languages spoken by at least $s$ people
thus should be quite flat for small $s$ instead of diverging
with a power law for $s \rightarrow 0$, as fitted in
\cite{brazil}. A log-normal distribution gives a much better 
overall fit and is for large sizes not necessarily worse than 
the two power laws of \cite{brazil}. 

Further results from the Viviane model are given in our appendix.

\subsubsection{Bit-string model}

Our own model uses bit-strings as in \cite{kosmidis,schwammle} but for 
different purposes. Each different \index{bit-string} bit-string
represents a different 
language though one may also define slightly different bit-strings as 
representing different dialects of the same language. Lengths $\ell$ of 
8 to 64 bits have been simulated, and the results for 16 bits differed little
from those of longer strings, while 8 bits behaved differently.

We used three different probabilities $p,\, q,\, r$ though most properties
can be also obtained form the special cases $q = 0, \; r = 1$. When a 
new individual is born its language is mutated with probability $p$ compared
to that of the mother. One of the $\ell$ bits is selected randomly and reverted,
which means a zero bit becomes one and a one bit becomes zero. This $p$ 
is the mutation probability per bit-string; the probability per bit is
therefore $p/\ell$. 

When $q$ is not zero, then the above mutation process is modified. With 
probability $1-q$ it happens randomly as above, and with probability $q$
the new value of the bit is obtained not by reverting it but by taking 
over the corresponding bit value of a randomly selected individual from
the whole population. This \index{transfer} transfer
probability $q$ thus describes the effect 
that one language can learn concepts from other languages. Many words of
higher civilization in the German language came from French, while French
beers sometimes have German names. 

Thus far the simulations are similar to biology with vertical ($p$) and 
horizontal ($q$) gene transfer. Specific human thinking enters into the third
probability $(1-x^2)r$ (also $(1-x)^2$ instead of $1-x^2$ was used)
to give up the own language and to switch to the 
language of another randomly selected person. Here $x$ is the fraction of
people speaking the old language, and thus this probability to abandon the
old language is particularly high for small languages. The new language is
selected by a random process, but since it is that of a randomly selected
person and not a randomly selected language, most likely the new language
is one of the major languages in the population. In this way we simulate
the same trend towards dominating language which was already modelled
by Abrams and Strogatz, as described above in the example of our emigration
to Brazil. This \index{flight} flight
from small to large languages, through the parameter $r$,
distinguishes the language competition from biological competition between
species in an ecosystem, and takes into account human consideration of 
the utility of the language. 

The population size is kept from going to infinity by a \index{Verhulst} 
Verhulst death probability proportional to the actual population size. Thus if 
we start with one person, the population will grow until it reaches the carrying
capacity given by the reciprocal proportionality factor. More practical is
an initial population which is already about equal to the final equilibrium 
population. With the latter choice one can start with either everybody
talking the same language, or everybody talking a randomly selected language.
A complete \index{Fortran} Fortran
program for the simple case $q=0, \; r=1$ is listed in \cite{newbook}.

Compared to the Viviane model explained above, our model is more complicated
since it has three probabilities $p,q,r$ instead of only one coefficient 
$\alpha$. However, one can set $q=0, \; r=1$ in our model and then has
the same number of free parameters. The Viviane model simulates the 
flight from small to large languages by a mutation probability inversely 
proportional to the size of the languages while we separate the mutations
(independent of language size) from the flight probability $(1 - x^2)r$. 
Moreover, we simulate a continuous competition of languages while the
Viviane model simulates the unique historical event of a human population
spreading over a continent where no humans lived before. 

The results of our model are reported in \cite{newbook,schulze}. Most important
is the sharp \index{phase transition} phase transition, for increasing mutation 
rate $p$ at fixed $q$ and $r$, between \index{dominance} dominance at small and 
\index{fragmentation} fragmentation at large $p$. For dominance,
at least three quarters of the population speak one language, and most of the 
others speak a variant differing by only one bit from that language. For
fragmentation, on the other hand, the population spreads over all possible
languages. If we start with dominance, the phase transition to fragmentation
was already described in the biblical story of the Tower of Babel. If we 
start with fragmentation, we get dominance for long enough times and small
enough mutation rates, if we 
use $(1-x)^2$  instead of $1-x^2$ for the flight probability. Fig.6 shows
the phase diagram for $\ell = 8$ and 16 if we start from fragmentation. In 
Fig.7, particularly long simulations for $\ell = 64$ and one million  people
show how an initial dominance decays into fragmentation. 

\begin{figure}[hbt]
\begin{center}
\includegraphics[angle=-90,scale=0.5]{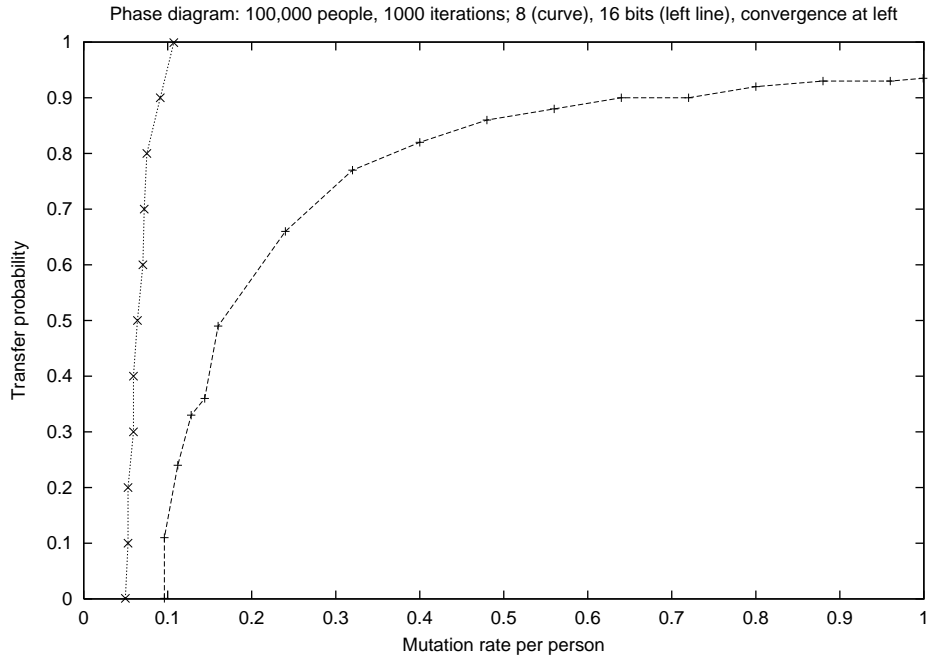}
\end{center}
\caption{Phase diagram for dominance in the upper left part and fragmentation
in the lower right part. The higher the mutation rate $p$ and the lower the
transfer rate $q$ is the more fragmented is the population into many 
different languages. We start with an equilibrium distribution of 100,000
languages. each speaking a randomly selected language. The curve corresponds
to $\ell = 8$ bits, the nearly straight line to $\ell = 16$; $r = 1$ in both
cases.  From \cite{schulze}.
}
\end{figure}

\begin{figure}[hbt]
\begin{center}
\includegraphics[angle=-90,scale=0.5]{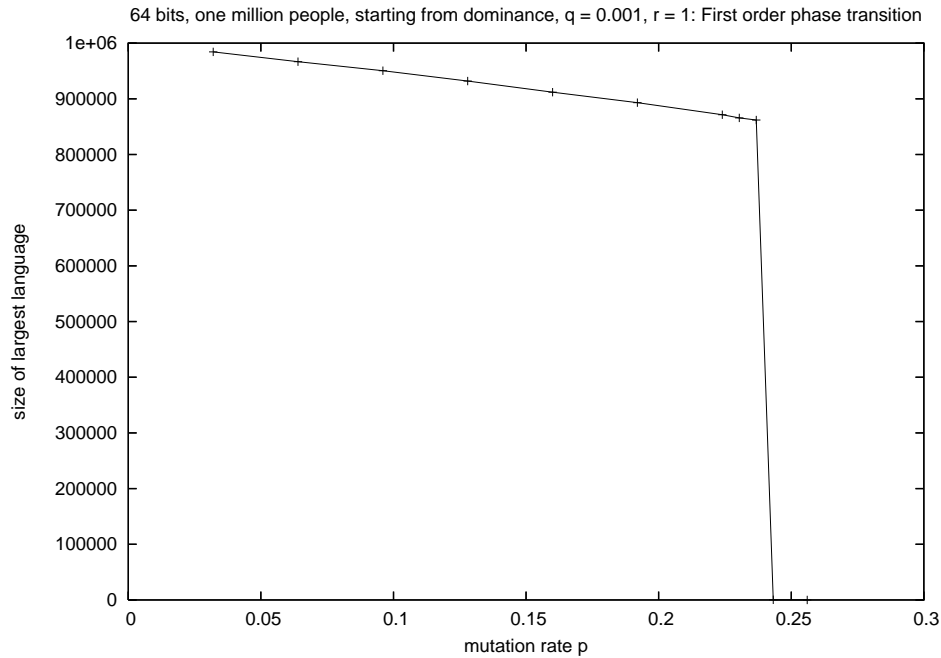}
\end{center}
\caption{Phase transition from dominance to fragmentation for one million
people and 64 bits, i.e. much more possible languages than people. We show
the size of the most-often spoken language after 300 iterations; it jumps from 
 $10^6 $ to $10^2$. 
}
\end{figure}

\begin{figure}[hbt]
\begin{center}
\includegraphics[angle=-90,scale=0.28]{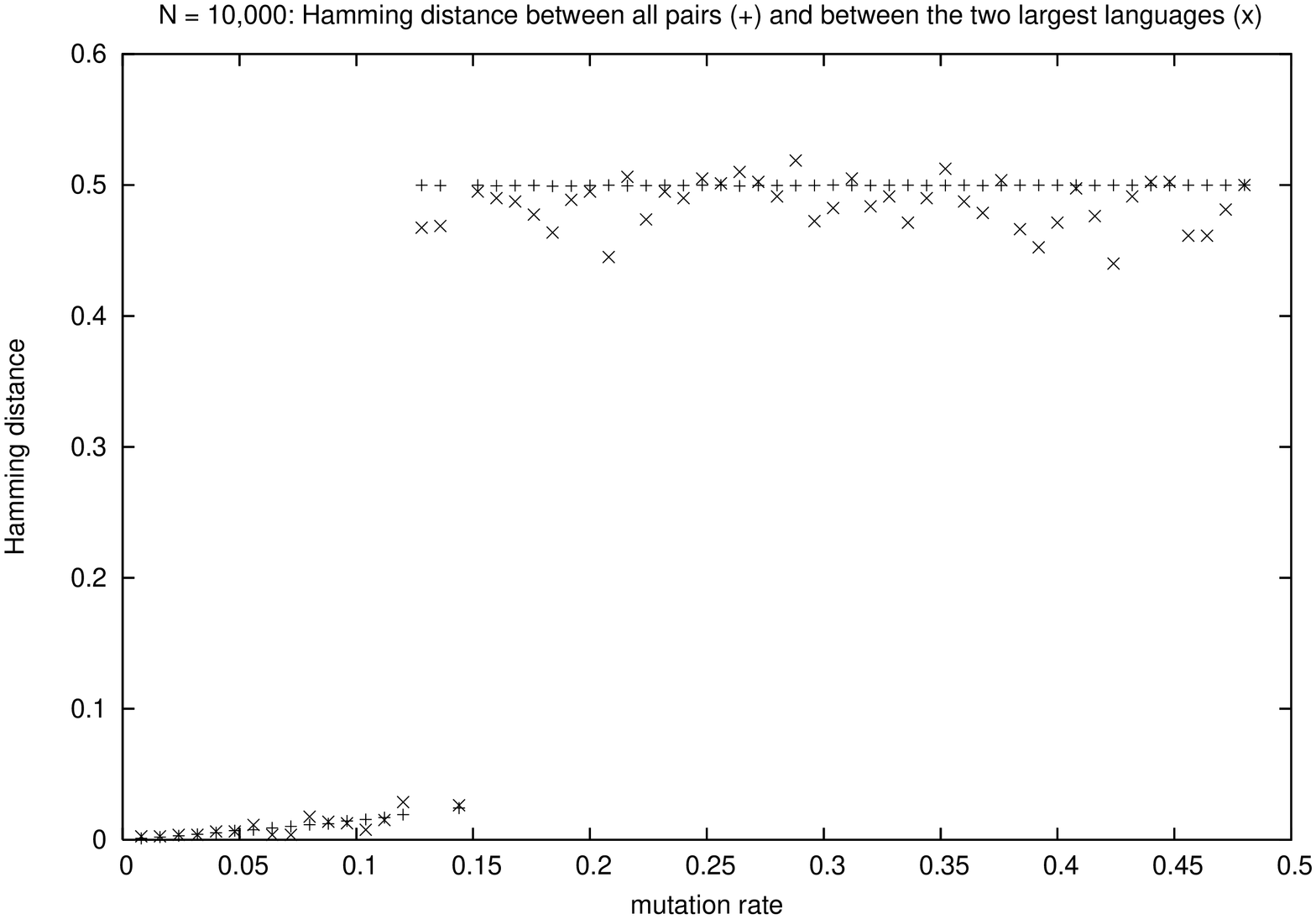}
\includegraphics[angle=-90,scale=0.28]{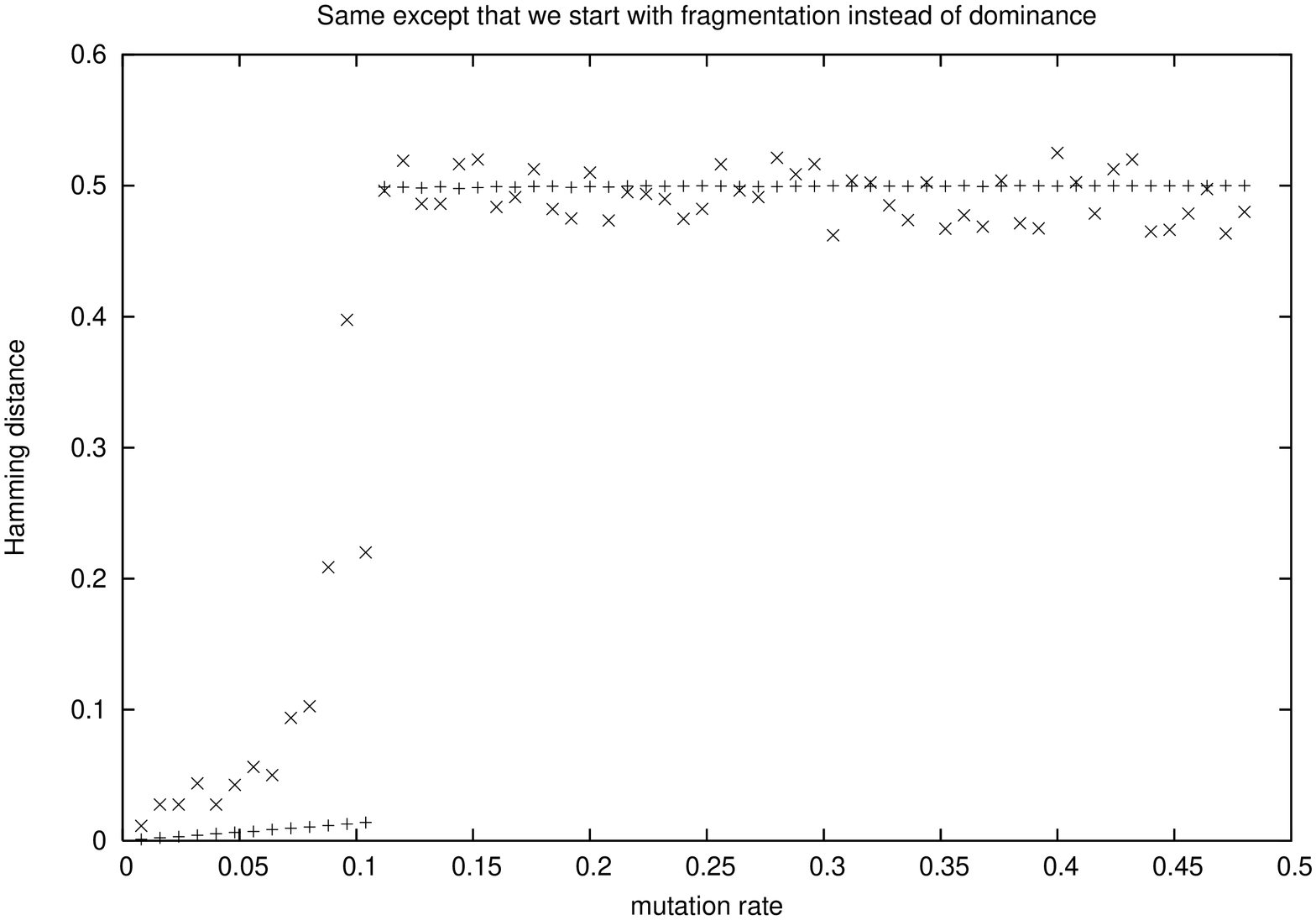}
\end{center}
\caption{Difference between languages, as measured by the normalized
Hamming distance = fraction of different bits. We show both the average
distance between all pairs and that between the two largest languages,
for 10,000 people and $q=0$. The top part starts with dominance, the
bottom part with fragmentation. From F.W.S. Lima, priv. comm.  
}
\end{figure}

\begin{figure}[hbt]
\begin{center}
\includegraphics[angle=-90,width=.72\textwidth]{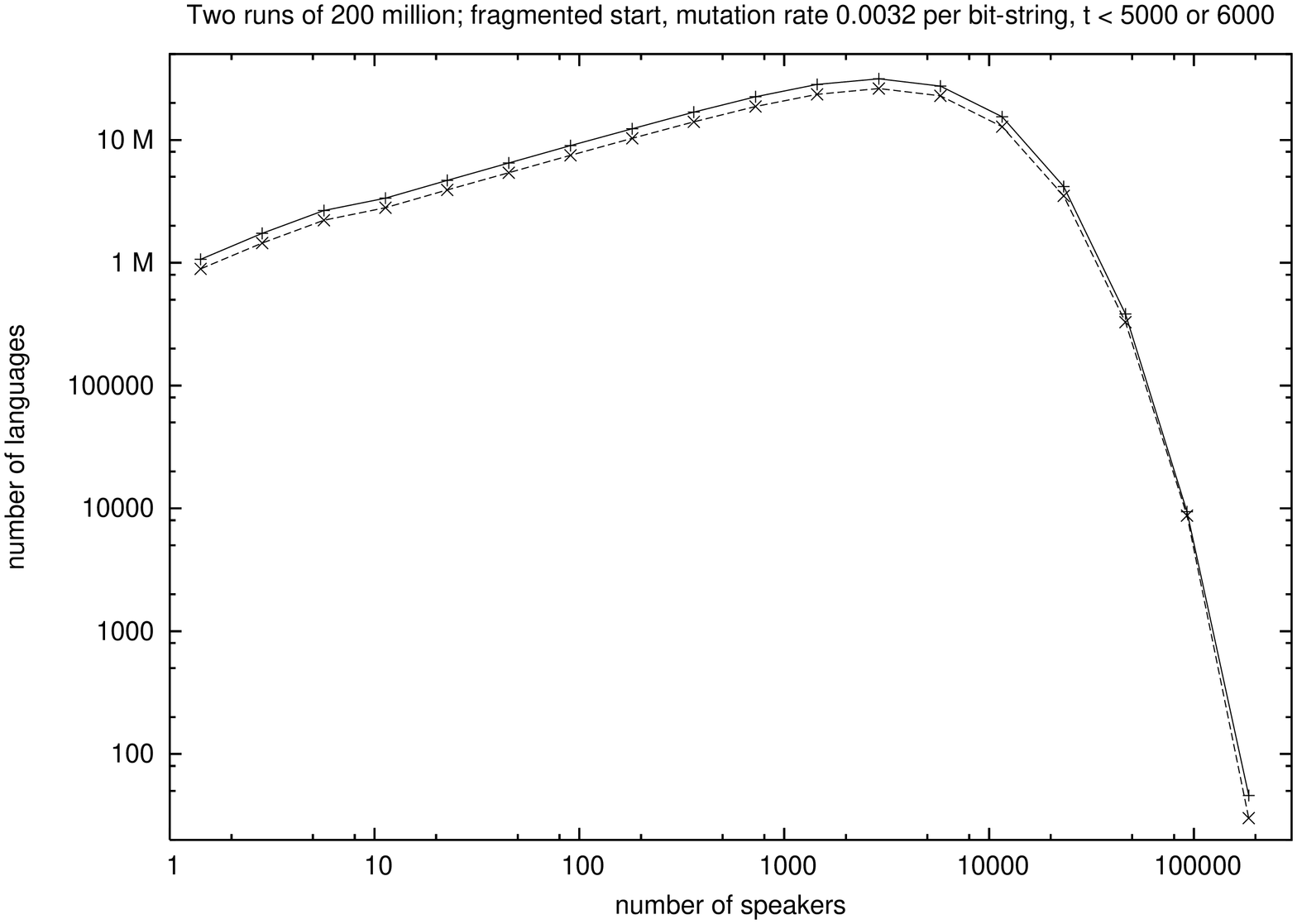}
\end{center}
\caption{
Size distribution far from equilibrium during the phase transiton
from fragmentation to dominance, $\ell = 16, \; q=0$. Additional 
smoothening by random multiplications was applied \cite{wichmann}.  
}
\end{figure}

Tesileanu and Meyer-Ortmanns \cite{tibihmo} introduced into this model
the \index{Hamming} Hamming
distance as a measure of dissimilarity between languages. This Hamming
distance counts the number of different bits in a position-by-position
comparison of two bit-strings. Thus the $\ell=4$ strings 0101 and 1010 have a
Hamming distance of four. This distance can be normalized to lie
between zero and one, through division by $\ell$. Fig.8 shows this normalized
Hamming distance for both the two largest languages and the average
over all possible pairs. Not much difference is seen except that the
one for the single pair fluctuates much stronger than the average over
all pairs. And for dominance the difference is very small while for
fragmentation is it nearly 1/2. Thus for fragmented populations, the
various languages are nearly uncorrelated, and half their bits agree 
accidentally while the other half disagrees. For dominance, the minor
languages are mostly one-bit mutants of the dominating language. Fig.8, like 
Fig.7 before, shows a clear first-order \index{phase transition} phase 
transition, that means a sharp jump. Thus far we were not able to
modify this model such that it gives a second-order transition where
the fraction of people speaking the largest language goes
continuously to zero at a sharp critical point. Such a modification
might give a more realistic distribution of language sizes.

The time dependence of the size of the largest cluster, if we start
with fragmentation, suggests a complicated nucleation
process. Originally all languages are about equal in size, and then
due to random fluctuations one language happens to be somewhat more
equal than the others. This language then wins over, first slowly,
then rapidly. The time needed for one language to win increases
about logarithmically when the population increases from $10^3$ to
$10^8$. Thus for an infinite population, as simulated by
deterministic differential equations of the Nowak et al style
\cite{nowak}, the emergence of dominance out of a fragmented
population might never happen in our model.  
 
First-order phase transitions like those in Figs. 7 and 8 are usually 
accompanied by hysteresis, like when undercooled liquid water is to crystallize
into ice. Thus we should get different positions of the effective
transition (for fixed population size and fixed observation time)
depending on whether we start from dominance or fragmentation. This is
shown in \cite{schulze}(b), using in both cases $(1-x)^2$ for the flight
probability. 

The language size distribution shows the desired shape of a
slightly asymmetric parabola on the log-log plot (log-normal distribution)
but the actual language sizes are far too small compared with
reality. This is not due to lack of computer power but comes from the 
sharp first-order transition, Figs.7 and 8. Either one language
dominates as if 80 percent of the world speaks Chinese. Or all $2^\ell$
languages are equivalent apart from fluctuations and thus each is
spoken only by a small population. If the first-order transition would
be changed into a second-order one, the results for mutation rates
slightly below the critical point might be better.

An alternative was suggested by linguist Wichmann \cite{wichmann}: The
present language distribution is not in equilibrium. If we assume that
parts of the world are on one side and parts on the other side of the 
phase transition from dominance to fragmentation (or from fragmentation
to dominance), then the above equilibrium results are not
good. Instead, we show in Fig. 9 two runs for a \index{non-equilibrium}
non-equilibrium
situation of about 5000 iterations at very low mutation rate, starting
from fragmentation. The results are averaged over the second half of
the simulation with the time adjusted such that the phase transition
of fig.11 happened during that second half. Now the language sizes
vary over five orders of magnitude, much better than before. (If
we start from dominance the size distribution is similar but more
symmetric \cite{wichmann}.) 

\section{Conclusion}

The last few years have seen the development of a variety of 
different approaches to simulate the competition between existing
languages of adult humans. Each model has its advantages and 
disadvantages.

If we follow the tradition of physics, that theories should 
explain precise experimental data, then the size histogram 
of the $10^4$ human languages, Fig.3, seems to be best candidate.
Empirically it is based on Grimes \cite{grimes} and was 
analyzed e.g. by \cite{sutherland,grimes,novotny,brazil}.
In order to simulate this language size distribution, we need
models for $10^4$ different languages, and only two of them
have been published thus far, the Viviane model and our model
\cite{schulze,gomes}.

Future work with these models could look at the similarities and 
differences between the languages (bit-strings), as started in 
\cite{tibihmo} and Fig.8, or the \index{geography} geography of languages and
their dialects \cite{goebl}, as started in \cite{gomes}.

We thank our coauthors \cite{wichmann,mallorca} for collaboration.

\section{Appendix}

This appendix brings some more results for many languages, first on
the Viviane model \cite{gomes} and then on our model \cite{schulze}.

\subsection{Viviane colonization model}
For the model of \cite{gomes}, one can look at the \index{history} history how
languages split from a mother languages, and later produced more
daughter languages. In contrast to linguistic field research, which looks 
only at the
last few thousand years, computer simulations can store and analyze
the whole history since the beginning. \cite{schulze}(d) shows for a small $64
\times 64$ lattice, how one language split into
daughters, etc, very similar to biological speciation trees. For
clarity we omitted numerous languages which had no "children". For 
larger lattices we found that even for many thousand languages a few 
steps suffice on average to reach from any of the languages in this
\index{tree} tree the oldest ancestor language on the top of the tree. Other tree 
simulations were published in \cite{wang}.  
   
Often a conquering population imposes its language to the native
population. Perhaps in Europe, before the arrival of Indo-European
farmers, the Cro Magnon people spoke a language family  of which the Basque
language is the only present survivor. Better documented, though not
necessarily more true, is the story of the single Gallic village in
today's France which resisted the Roman conquest two millennia ago, thanks to 
the efforts of \index{Asterix} Asterix and Obelix (helped by doping). In the
Viviane model, where people may adopt the language of their
neighbours, such a single resistance center can influence many other
sites during the later spread of languages. indeed a
rather large fraction of the total population is influenced by
Asterix, particularly for large mutation rates \cite{schulze}(d). In 
physics, such simulation of the influence of a single "error" are
called \index{damage} "damage spreading".

\begin{figure}[hbt]
\begin{center}
\includegraphics[angle=-90,scale=0.27]{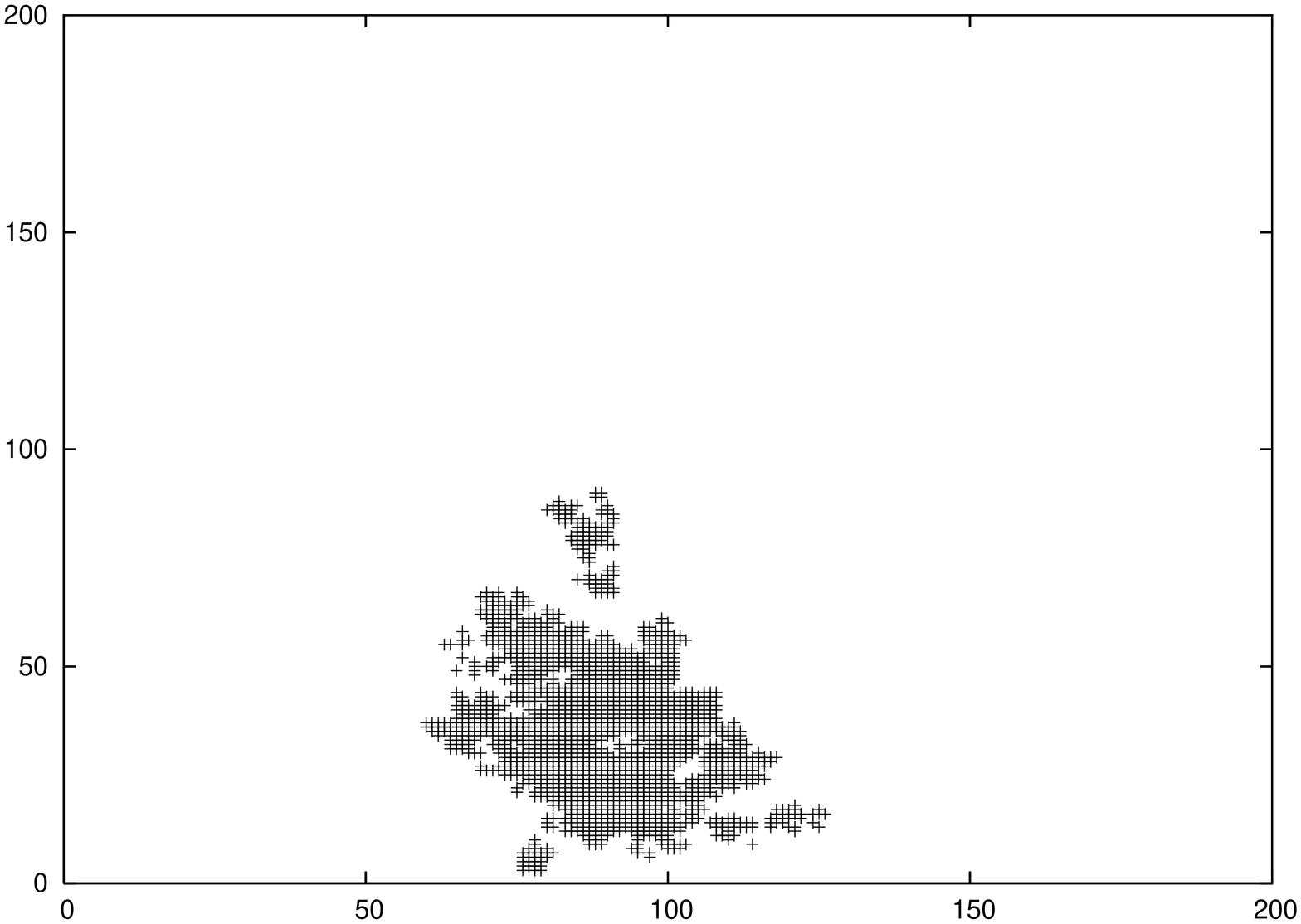}
\includegraphics[angle=-90,scale=0.27]{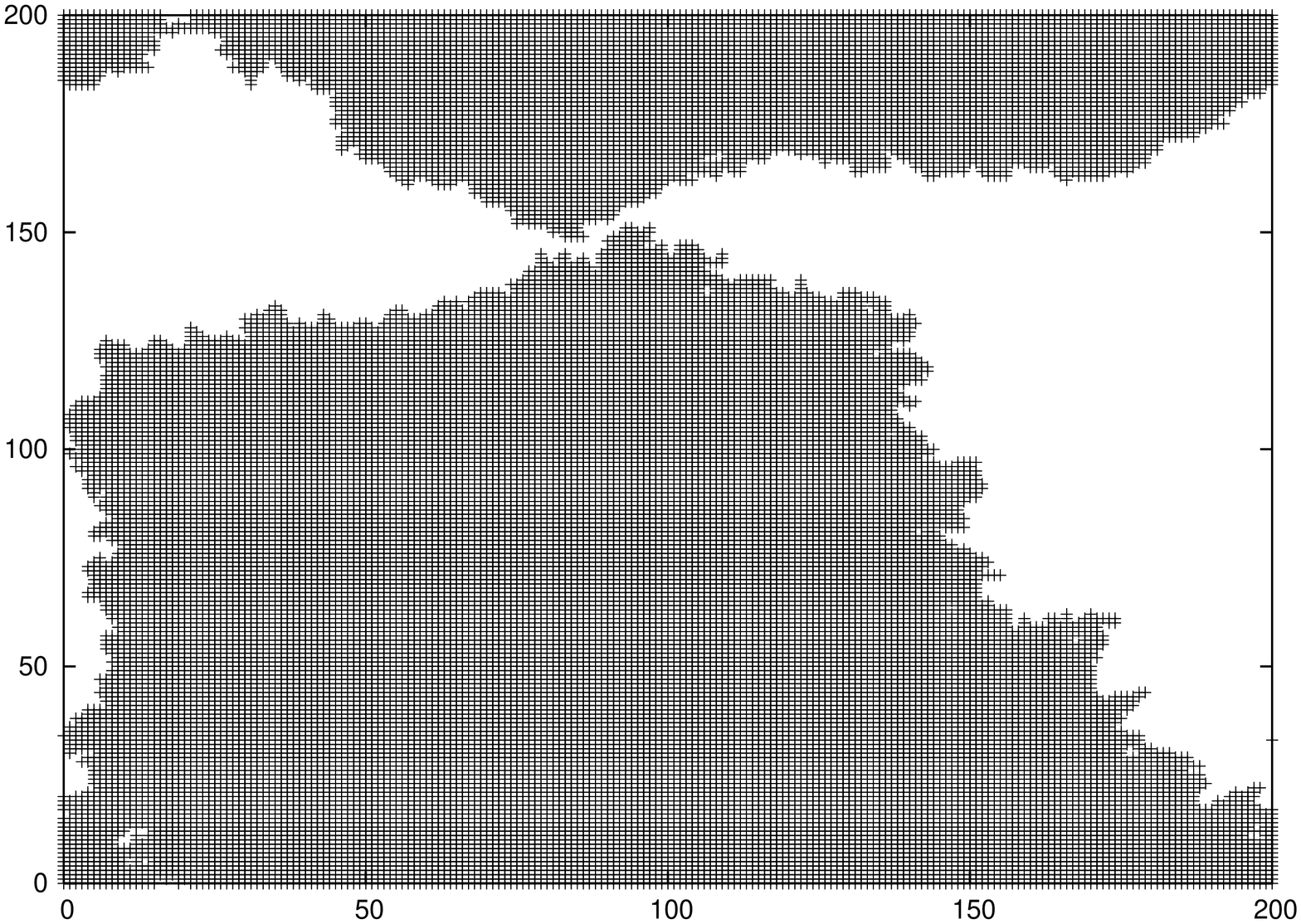}
\end{center}
\caption{
Domain formation if flight and transfer happen only to/from a language learned 
from a
lattice neighbour. We mark the sites where the largest language is spoken, after
240 and 450 iterations. For $t \ge 514$ nearly everybody speaks this
language. $(L = 200,\; p = 0.016, \; q = 0.9, \; r=1, \; \ell=16, \; Q=2,$ 
periodic boundary conditions). 
}
\end{figure}

\begin{figure}[hbt]
\begin{center}
\includegraphics[angle=-90,scale=0.3]{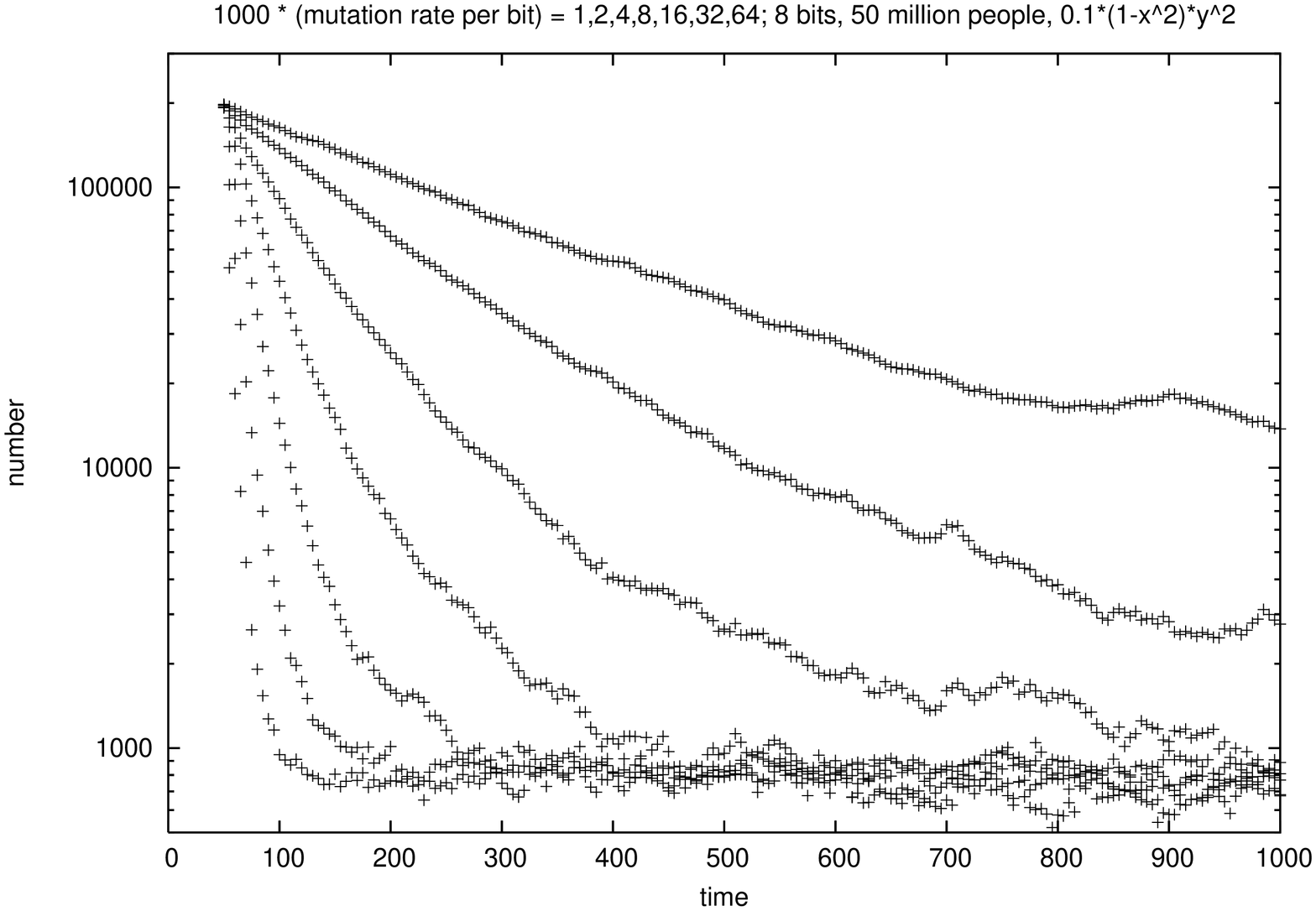}
\includegraphics[angle=-90,scale=0.3]{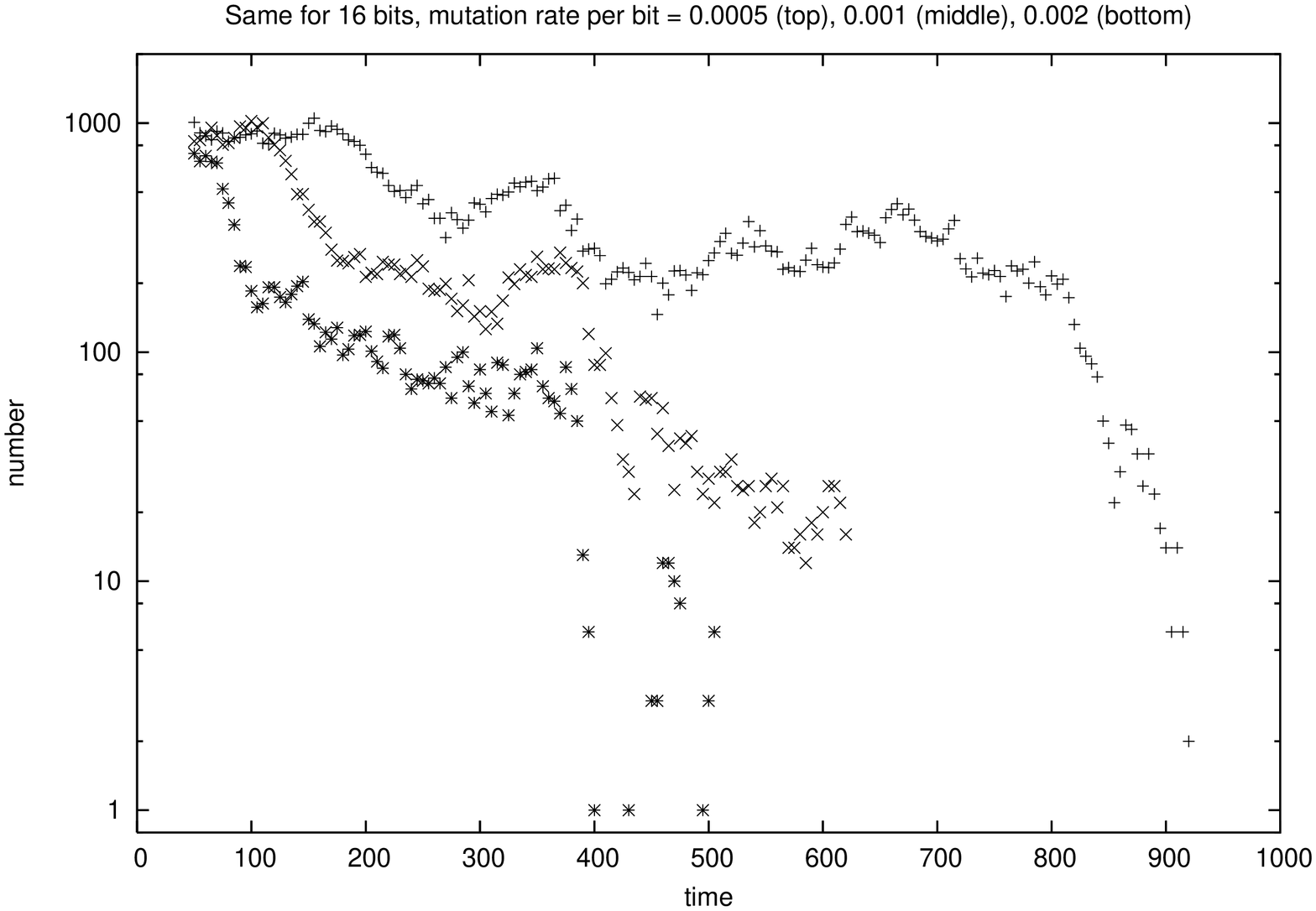}
\end{center}
\caption{
Decay of population originally speaking one particular language for 
$\ell = 8$ (top) and 16 (bottom). 
}
\end{figure}

\begin{figure}[hbt]
\begin{center}
\includegraphics[angle=-90,width=.72\textwidth]{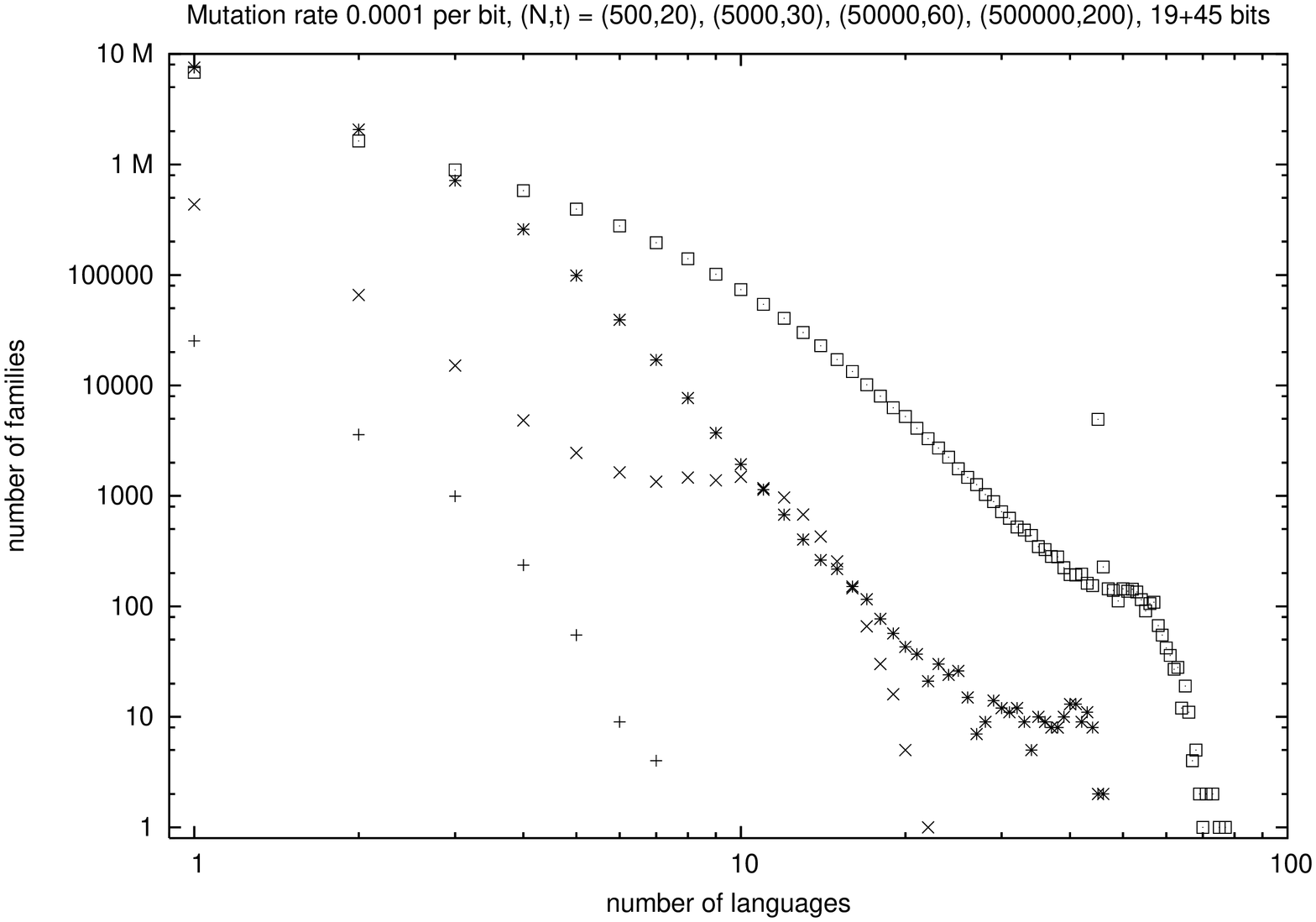}
\end{center}
\caption{
Distribution of the number of languages in a language family,
from sums over 100 or 10 independent simulations at various 
population sizes, $p = 0.0064$. The observation time increases slightly with 
increasing population size.
From ongoing work with S. Wichmann and F.W.S. Lima.
}
\end{figure}

\subsection{Our bit-string model}
While the Viviane model always happens on a lattice \index{lattice}, for our model the
lattice is optional. If we want to study the geographical coexistence
of two languages in adjacent regions, then of course a lattice is
needed \cite{schulze}(b). Now on every lattice site live many people. 
Without any difference in status, as opposed to
\cite{finland}, on one side one language dominates and on the other
side the other language, if initially each region was occupied
only by speakers of its own language. Also in the transition
region the other $2^\ell-2$ languages play no major role. The situation in this 
figure may correspond to \index{New Orleans} New Orleans a long time
ago, where Canal Street separated the French quarter from the newer 
English settlement. These methods could be applied to dialectometry,
\index{dialectometry} as documented for France by Goebl \cite{goebl}.

Bit-strings allow only $Q=2$ choices per position, but the lattice model was
also generalized to $Q = 3$ and 5 choices. Surprisingly, the phase 
transition curve \cite{schulze}(d) between dominance for low and fragmentation
for high mutation rates was independent of this number $Q$ of choices. Only
when $\ell$ was changed, the different transition curves were
obtained. 

If we want to apply the lattice model to \index{geography} geography we want 
compact
geographical language regions to emerge from a fragmented start. Then not
only the transfer of language elements but also the flight to another 
language needs to be restricted to lattice neighbours, i.e. people learn
new elements or a new language only from one of the four nearest
neighbours, randomly selected. Fig.10 shows how one language, accidentally 
the largest at intermediate times, grows until
is covers nearly the whole lattice.

One may look, without lattice, on the \index{history} history of people
speaking one randomly selected language in an initially fragmented 
population. Because of mutations, after a long enough time everybody
has moved at least once to another language. But since the number 
$L = 2^\ell$ of possible languages is finite, some people move back to
their original language, like emigrants whose offspring later return
to their old country. Thus after 50 iterations to
give an equilibrium, we mark all those speaking language zero. Their
offspring carries that mark also, even if they mutate their
language, and we count at each time step the number of marked people 
speaking language zero. 

Then we see a rapid decrease of that number; to slow down the decay we 
modified the flight probability to $0.1(1-x^2)y^2$ where $y$ is the
fraction for the language which the individual from fraction $x$ 
considers to switch to. Then a slower decay as in Fig.11 results,
faster for higher mutation rates. For $\ell = 8$ bits we see nicely
the random background of less than thousand people (from 50 million) 
who returned to the language zero of their ancestors; for $\ell = 16$ 
both the initial and the final number of zero speakers are much
smaller since the 50 million can now distribute among 65536 instead of
only 256 possible languages. 

Human languages can be grouped into families, like the Indo-European family
\index{family} of $10^2$ different languages. To simulate language families
we need a criterion which bit-strings belong into one family. 
Thus we worked with $\ell = 64$ bits and assumed, following 
Wichmann, that the leading 19 bits determine the family and the
remaining 45 bits the different languages within one family. The 
numbers $2^{19}$ and $2^{45}$ of possible families and languages
are so large that our computer simulations with less than a 
million people do not notice their finite size. Indeed, the
results in Fig.12 for 500,\, 5000,\, 50,000 and 500,000 people are
roughly independent of population size and show a mostly monotonically
decaying probability distribution function for the number of
languages within one family. Empirical observations were 
published e.g. by Wichmann \cite{grimes}.

\def\bibindent{6mm}

\bigskip 
This is a shortened version of a chapter for 
Econophysics \& Sociophysics: Trends \& Perspectives, edited by
B.K. Chakrabarti, A. Chakraborti, A. Chatterjee for WILEY-VCH, Weinheim 2006.
\end{document}